\documentclass[twocolumn]{revtex4}
\usepackage{epsfig}
\usepackage{amssymb}
\usepackage{amsmath}
\usepackage{amsfonts}

\newcommand{\R}{\mathbb{R}}
\newcommand{\C}{\mathbb{C}}

\newcommand{\fb}{\mathfrak{b}}

\newcommand{\fz}{\mathfrak{z}}

\newcommand{\fB}{\mathfrak{B}}

\newcommand{\be}{\begin{equation}}
\newcommand{\ee}{\end{equation}}
\newcommand{\bea}{\begin{eqnarray}}
\newcommand{\eea}{\end{eqnarray}}

\newcommand{\kt}{\rangle}
\newcommand{\br}{\langle}

\newcommand{\ed}{\end{document}}

\newcommand{\pbr}{\prec\!}
\newcommand{\pkt}{\!\succ}

\newcommand{\cH}{\mathcal{H}}

\newcommand{\pH}{{\cal H}_{\rm phys}}

\newcommand{\cP}{{\cal P}}

\newcommand{\cT}{\mathcal{T}}

\newcommand{\tr}{{\rm tr}}
\begin{document}
\title{On Hamiltonians Generating Optimal-Speed Evolutions}

\author{Ali~Mostafazadeh}
\address{Department of Mathematics, Ko\c{c} University, Sariyer 34450,
Istanbul, Turkey\\
amostafazadeh@ku.edu.tr}

\begin{abstract}
We present a simple derivation of the formula for the Hamiltonian
operator(s) that achieve the fastest possible unitary evolution
between given initial and final states. We discuss how this formula
is modified in pseudo-Hermitian quantum mechanics and provide an
explicit expression for the most general optimal-speed
quasi-Hermitian Hamiltonian. Our approach allows for an explicit
description of the metric- (inner product-) dependence of the lower
bound on the travel time and the universality (metric-independence)
of the upper bound on the speed of unitary evolutions.
\medskip

\hspace{6.1cm}{Pacs numbers: 03.65.Xp, 03.67.Lx, 02.30.Yy, 02.40.-k}
\end{abstract}
%\pacs{Pacs numbers: 41.20.Jb, 42.25.Bs}

\maketitle

In quantum mechanics the travel time for unitary time-evolutions
between an initial and a final states $\lambda_I$ and $\lambda_F$
has a minimum that is proportional to the distance between
$\lambda_I$ and $\lambda_F$ in the state space
\cite{anandan-aharonov}. For a system with a fixed energy scale,
this implies that the speed of unitary evolutions has an upper
bound. The problem of determining a Hamiltonian operator that
achieves the highest evolution speed has been considered in
\cite{fleming-Vaidman,Margolus,brody,carlini,brody-hook}. The
purpose of the present article is to offer a very simple derivation
of the formula for a time-independent optimal-speed Hamiltonian that
can be directly generalized to the pseudo-Hermitian representation
of quantum mechanics \cite{jpa-2004b}. In particular, we give, for
the first time, the explicit form of the most general
time-independent quasi-Hermitian \cite{quasi} optimal-speed
Hamiltonian that evolves $\lambda_I$ into $\lambda_F$.

In the standard formulation of quantum mechanics, the (pure) states
of a physical system are identified with the rays in a complex
separable Hilbert space $\cH$. Each ray can be determined in terms
of an associated state vector $\psi\in\cH-\{0\}$ according to
$\lambda_\psi=\{c\psi|c\in\C\}$. It is usually convenient to use the
one-to-one correspondence between the states $\lambda_\psi$ and the
projection operators
$\Lambda_\psi:=\frac{|\psi\kt\br\psi|}{\br\psi|\psi\kt}$ to describe
the properties of the space of all states, i.e., the projective
Hilbert space $\cP(\cH)$. For a Hilbert space $\cH$ of dimension
$N\leq\infty$, $\cP(\cH)$ is the complex projective space $\C
P^{N-1}$ that plays a central role in the description of geometric
phases \cite{page}.

It is an easy exercise to show that $\Lambda_\psi$ satisfies
    \be
    \Lambda_\psi^2=\Lambda_\psi=\Lambda_\psi^\dagger,~~~~~~~
    \tr(\Lambda_\psi)=1,
    \label{id1}
    \ee
where ``$\tr$'' denotes the trace. Recall that for a linear operator
$L$ acting in $\cH$, $\tr(L):=\sum_{n=1}^N\br\xi_n|L\,\xi_n\kt$,
where $\{\xi_n\}$ is an arbitrary orthonormal basis of
$\cH$,\footnote{$\tr(L)$ is independent of the choice of
$\{\xi_n\}$.}.

In view of (\ref{id1}), $\Lambda_\psi$ is an element of the space
$\fB_2(\cH)$ of all linear operators $L:\cH\to\cH$ fulfilling
$\tr(L^\dagger L)<\infty$. We can use ``$\tr$'' to define the
following inner product on this space $(L|J):=\tr(L^\dagger J)$.
This makes $\fB_2(\cH)$ into a separable Hilbert space known as the
Hilbert-Schmidt class \cite{reed-simon}. Because the state space
$\cP(\cH)$ is a subset of $\fB_2(\cH)$, we can use the inner product
$(\cdot|\cdot)$ to define a notion of distance (metric) on
$\cP(\cH)$. We define the line element $ds$ on $\cP(\cH)$ according
to
    \be
    ds^2:=\frac{1}{2}\,(d\Lambda_\psi|d\Lambda_\psi)=
    \frac{\br\psi|\psi\kt\br d\psi|d\psi\kt-|
    \br\psi|d\psi\kt|^2}{\br\psi|\psi\kt^2},
    \label{Fubini-Study=}
    \ee
where we have used
$\Lambda_\psi:=\frac{|\psi\kt\br\psi|}{\br\psi|\psi\kt}$ and
(\ref{id1}), \cite{prl-2007}. For $N<\infty$ where $\psi$ can be
represented by a complex column vector $\vec\fz$ with components
$\fz_1,\fz_2,\cdots,\fz_N$, (\ref{Fubini-Study=}) takes the form
$ds^2=\sum_{a,b=1}^N g_{ab^*}d\fz_ad\fz_b^*$ where $
g_{ab^*}:=|\vec\fz|^{-4}(|\vec\fz|^2\delta_{ab}-\fz_a^*\fz_b)$. This
is precisely the Fubini-Study metric tensor
\cite{eguchi-gilky-hanson}. For $N=2$, endowing $\cP(\cH)$ with this
metric yields a round two-dimensional sphere of unit diameter.

Now, suppose that we wish to use an arbitrary Hermitian
(self-adjoint) Hamiltonian operator $H:\cH\to\cH$ to evolve an
initial state $\lambda_{\psi_I}$ to a final state
$\lambda_{\psi_F}$. We can view the evolving state
$\lambda_{\psi(t)}$ as a point moving on $\cP(\cH)$. According to
(\ref{Fubini-Study=}), the instantaneous speed of the evolution is
given by
    \[
    \frac{ds}{dt}=\mbox{\large$
    \frac{\sqrt{\br\psi(t)|\psi(t)\kt\br\dot\psi(t)|\dot\psi(t)\kt-
    |\br\psi(t)|\dot\psi(t)\kt|^2}}{\br\psi(t)|\psi(t)\kt}$}=
    \frac{\Delta E_{\psi(t)}}{\hbar},\]
where
    \be
    \Delta E_{\psi(t)}:=\sqrt{
    \frac{\br\psi(t)|H^2\psi(t)\kt}{\br\psi(t)|\psi(t)\kt}
    -\frac{|\br\psi(t)|H\psi(t)\kt|^2}{\br\psi(t)|\psi(t)\kt^2}},
    \label{uncertainty=}
    \ee
is the uncertainty in energy and we have employed the Schr\"odinger
equation, $H\psi(t)=i\hbar\dot\psi(t)$. We can integrate $ds/dt$ to
obtain the length of the curve traced by $\lambda_{\psi(t)}$ in
$\cP(\cH)$ as a function of the travel time $\tau$,
\cite{anandan-aharonov}:
    \be
    s=\frac{1}{\hbar}\int_0^\tau \Delta E_{\psi(t)}\,dt.
    \label{length}
    \ee
Because $\Delta E_{\psi(t)}\geq 0$ for all $t\in[0,\tau)$, $s$ is a
monotonically increasing function of $\tau$. This makes $\tau$ a
monotonically increasing function of $s$. Therefore, the shortest
travel time is achieved for the paths of the shortest length, i.e.,
the geodesics on $\cP(\cH)$, \cite{anandan-aharonov}.

Note that the geodesic distance is uniquely determined by the
initial and final states and is insensitive to the choice of the
Hamiltonian one uses to evolve the initial state along such a
geodesic. The travel time depends on the Hamiltonian through the
energy uncertainty $\Delta E_{\psi(t)}$. In particular, if one can
make the latter arbitrarily large, the travel time can be made
arbitrarily small. In typical situations, however, $\Delta
E_{\psi(t)}$ has a constant upper bound. For example, consider the
case that the Hilbert space is finite-dimensional ($N<\infty$) and
the energy eigenvalues $E_n$ are bounded functions of time; there is
some ${\cal E}\in\R^+$ such $|E_n(t)|\leq {\cal E}$ for all $n$ and
$t$. Then, we can easily show that $\Delta E_{\psi(t)}\leq {\cal
E}$; the travel speed is bounded by ${\cal E}/\hbar$; and the travel
time has $\hbar s/{\cal E}$ as a lower bound. Here $s$ is to be
identified with the geodesic distance between the initial and final
states.

The above argument is valid, if one does not have additional
restrictions on the choice of the Hamiltonian. In practice, one may
have to impose constraints that would make it impossible to evolve
the initial state along the shortest geodesic connecting it to the
final state. In this case one can formulate the problem as a
constrained variational problem \cite{carlini}. In the remainder of
this article we consider constant unconstrained Hamiltonians where
the minimum travel time depends, besides the geodesic distance
between the initial and final states, on a single real parameter
specifying the energy scale of the system.

Let $H$ be a time-independent Hamiltonian operator. Then the
time-evolution operator $e^{-itH/\hbar}$ commutes with $H$ and
$H^2$, and $\Delta E_{\psi(t)}$ does not depend on $t$. In this
case, (\ref{length}) implies $\tau=\hbar s/\Delta E_{\psi}$, and the
speed of the evolution is given by $\Delta E_{\psi}/\hbar$.
Therefore, to achieve the highest speed we need to choose the
Hamiltonian so that $\Delta E_{\psi}/\hbar$ is maximized. This shows
that the travel time is bounded by the ratio of the minimum of $s$,
i.e., the geodesic distance between $\lambda_{\psi_I}$ and
$\lambda_{\psi_F}$, to the maximum of speed $\Delta E_{\psi}/\hbar$.

Because we require the evolving state $\lambda_{\psi(t)}$ to trace a
geodesic in $\cP(\cH)$ that connects $\lambda_{\psi_I}$ and
$\lambda_{\psi_F}$, it lies entirely in the projective Hilbert space
$\cP(\cH')$ where $\cH'$ is the subspace of $\cH$ spanned by
$\psi_I$ and $\psi_F$. This is in fact a characteristic property of
the Fubini-Study metric \cite{anandan-aharonov,bz}. It shows that we
can restrict our attention to the case that $\cH$ is
two-dimensional; $N=2$, \cite{brody}. Furthermore, without loss of
generality, we can suppose that $\tr(H)=0$. This implies that the
eigenvalues of $H$ have opposite sign, $E_2=-E_1=:E$. Let
$\{\psi_1,\psi_2\}$ be an orthonormal basis consisting of the
eigenvectors of $H$, $H\psi_n=E_n\psi_n$. We expand $\psi(0)=\psi_I$
in this basis to find
    \be
    \psi_I=c_1\psi_1+c_2\psi_2,~~~~~c_1,c_2\in\C,
    \label{psi-expand-ini}
    \ee
and use the time-independence of $\Delta E_\psi$ to compute it at
$t=0$. In view of (\ref{uncertainty=}) and (\ref{psi-expand-ini}),
this yields
    \be
    \Delta E_\psi=E \sqrt{1-\left(\frac{|c_1|^2-|c_2|^2}{
    |c_1|^2+|c_2|^2}\right)^2}\leq E.
    \label{delta-E=}
    \ee
Therefore, the travel time $\tau$ satisfies
    \be
    \tau\geq \tau_{\rm min}:=\frac{\hbar s}{E},
    \label{bound}
    \ee
where $s$ is the geodesic distance between $\lambda_{\psi_I}$ and
$\lambda_{\psi_F}$ in $\cP(\cH)$. (\ref{bound}) identifies
$\tau_{\rm min}$ with a lower bound on the travel time. Next, we
construct a Hamiltonian $H_\star$ with eigenvalues $\pm E$ for which
$\tau=\tau_{\rm min}$. This shows that indeed $\tau_{\rm min}$ is
the minimum travel time.

Because $s$ is completely determined by $\lambda_{\psi_I}$ and
$\lambda_{\psi_F}$, the condition $\tau=\tau_{\rm min}$ is fulfilled
if and only if $\Delta E_\psi=E$. In light of (\ref{delta-E=}) this
is equivalent to $|c_1|=|c_2|$. If we expand $\psi_F$ in the basis
$\{\psi_1,\psi_2\}$ to find
    \be
    \psi_F=d_1\psi_1+d_2\psi_2,~~~~~d_1,d_2\in\C,
    \label{psi-expand-fin}
    \ee
and compute $\Delta E_\psi$ at $t=\tau$, we obtain (\ref{delta-E=})
with $(c_1,c_2)$ replaced with $(d_1,d_2)$. As a result, in order to
maintain $\Delta E_\psi=E$, we must have $|d_1|=|d_2|$.

Next we express $|c_1|=|c_2|$ and $|d_1|=|d_2|$ in the form
$c_2=e^{i\alpha_{_I}}c_1$ and $d_2=e^{i\alpha_{_F}}d_1$, for some
$\alpha_{_I},\alpha_{_F}\in\R$, respectively. Substituting these in
(\ref{psi-expand-ini}) and (\ref{psi-expand-fin}), we find
    \be
    \psi_1+e^{i\alpha_{_I}}\psi_2=c_1^{-1}\psi_I,~~~~~~
    \psi_1+e^{i\alpha_{_F}}\psi_2=d_1^{-1}\psi_F.
    \label{psi-eqns}
    \ee
We can solve these equations for $\psi_1$ and $\psi_2$ in terms of
$\psi_I$ and $\psi_F$, and use the spectral resolution of $H_\star$,
i.e.,
    \be
    H_\star=E\big(-|\psi_1\kt\br\psi_1|+|\psi_2\kt\br\psi_2|\big),
    \label{spec-res}
    \ee
to compute $H_\star$. This calculation is more conveniently
performed in terms of
    \be
    \vartheta:=\alpha_{_I}-\alpha_{_F},~~~~~~~
    \hat\psi_I:=\frac{\psi_I}{\sqrt 2\, c_1},~~~~~~~
    \hat\psi_F:=\frac{e^{\frac{i\vartheta}{2}}\psi_F}{
    \sqrt 2\, d_1}.
    \label{vartheta=}
    \ee
The result is \cite{carlini,brody-hook}
    \bea
    H_\star&=&
    \frac{iE\big(|\hat\psi_F\kt\br\hat\psi_I|-
    |\hat\psi_I\kt\br\hat\psi_F|\big)}{4\sin(\frac{\vartheta}{2})}
    \label{H-star-old}\\
    &=&
    \frac{iE\cot(\mbox{$\frac{\vartheta}{2}$})}{4}
    \left(\frac{|\psi_F\kt\br\psi_I|}{\br\psi_I|\psi_F\kt}-
    \frac{|\psi_I\kt\br\psi_F|}{\br\psi_F|\psi_I\kt}\right),
    \label{H-star=expl}
    \eea
where we have used the fact that $\hat\psi_I$ and $\hat\psi_F$ are
unit vectors. Moreover, (\ref{vartheta=}) implies $
\cos^2(\mbox{$\vartheta/2$})=
\frac{|\br\psi_I|\psi_F\kt|^2}{\br\psi_I|\psi_I\kt\,
\br\psi_F|\psi_F\kt},$ which as explained in \cite{anandan-aharonov}
identifies $\vartheta$ with $2s$, \footnote{In view of this
expression for $\cos^2(\vartheta/2)$, (\ref{vartheta=}), and
$\br\hat\psi_I|\hat\psi_I\kt=\br\hat\psi_F|\hat\psi_F\kt=1$,
$c_1=\sqrt{\br\psi_I|\psi_I\kt/2}\,e^{i\gamma}$ and
$d_1=\sqrt{\br\psi_F|\psi_F\kt/2}\,e^{i\delta}$, where
$\gamma,\delta\in\R$ are arbitrary if $\br\psi_I|\psi_F\kt=0$,
otherwise $e^{i(\delta-\gamma)}=e^{\frac{i\vartheta}{2}}
\br\psi_I|\psi_F\kt/|\br\psi_I|\psi_F\kt|$.}.

Equation (\ref{H-star=expl}) can be easily modified to give the
expression for the optimal-speed Hamiltonians in pseudo-Hermitian
quantum mechanics. One merely needs to make the following
substitution in the above analysis
    \be
     |\psi_n\kt \to |\psi_n\pkt,~~~
    \br\psi_n| \to \:\pbr\psi_n|:=\br\psi_n|\eta_+,
    ~~~s\to s_{\eta_+},~~~
    \label{subs}
    \ee
where $\eta_+:\cH\to\cH$ is the metric operator that defines the
inner product of the physical Hilbert space $\pH$, i.e.,
$\br\cdot,\cdot\kt_{\eta_+}:=\br\cdot|\eta_+\cdot\kt=\pbr\cdot|\cdot\pkt$,
and $s_{\eta_+}$ is the distance defined by the natural metric on
the projective Hilbert space $\cP(\pH)$. We can obtain the line
element associated with this metric by making the substitutions
(\ref{subs}) on the right-hand side of (\ref{Fubini-Study=}). This
gives \cite{prl-2007}
    \be
    ds_{\eta_+}^2:=
    \frac{\pbr\psi|\psi\pkt\pbr d\psi|d\psi\pkt-|
    \pbr\psi|d\psi\pkt|^2}{\pbr\psi|\psi\pkt^2}.
    \label{ph-Fubini-Study=}
    \ee
Similarly we find the following expressions for minimum travel time
$\tau^{(\eta_+)}_{\rm min}$ and the optimal-speed
$\eta_+$-pseudo-Hermitian \cite{p1} Hamiltonian $H_\star^{(\eta_+)}$
(with eigenvalues $\pm E$).
    \bea
    \tau^{(\eta_+)}_{\rm min}&=&\frac{\hbar s_{\eta_+}}{E},
    \label{ph-min-time}\\
    H_\star^{(\eta_+)}&=&
    \frac{iE\cot(s_{\eta_+})}{4}
    \left(\frac{|\psi_F\pkt\pbr\psi_I|}{\pbr\psi_I|\psi_F\pkt}-
    \frac{|\psi_I\pkt\pbr\psi_F|}{\pbr\psi_F|\psi_I\pkt}\right)
    ,~~~~~~\label{ph-H-star}
    \eea
where
    \bea
    \cos^2(s_{\eta_+})&=&
    \frac{|\pbr\psi_I|\psi_F\pkt|^2}{\pbr\psi_I|\psi_I\pkt\,
    \pbr\psi_F|\psi_F\pkt}.
    \label{ph-cosvartheta=}
    \eea

According to (\ref{bound}) and (\ref{ph-min-time}), if we choose
$\eta_+$ such that the geodesic distance $s_{\eta_+}$ between
$\lambda_{\psi_I}$ and $\lambda_{\psi_F}$ in $\cP(\pH)$ is smaller
than the geodesic distance $s$ between $\lambda_{\psi_I}$ and
$\lambda_{\psi_F}$ in $\cP(\cH)$, then $\tau^{(\eta_+)}_{\rm
min}<\tau_{\rm min}$. This is the essence of the main result of
\cite{bbj-prl-2007}. Indeed, as we show below, it is possible to
choose $\eta_+$ so that regardless of the choice of
$\lambda_{\psi_I}$ and $\lambda_{\psi_F}$ their distance in
$\cP(\pH)$ becomes arbitrarily small. But this does not seem to have
any physical implications, for such an evolution amounts to evolving
a state to an arbitrarily close state in an arbitrarily short time.
The physical quantity of practical significance, particularly in
areas such as quantum computation, is the speed of the evolution,
namely $E/\hbar$, which is a universal quantity independent of the
choice of $\eta_+$. Therefore, the minimum travel time between
states of a given distance is independent of $\eta_+$,
\cite{prl-2007}. A physical process that involves evolving
$\lambda_{\psi_I}$ into $\lambda_{\psi_F}$ in $\cP(\pH)$ using an
$\eta_+$-pseudo-Hermitian Hamiltonian $H:\cH\to\cH$ in time $\tau$
may be described equally well by evolving
$\lambda_{\eta_+^{1/2}\psi_I}$ into $\lambda_{\eta_+^{1/2}\psi_F}$
in $\cP(\cH)$ using the equivalent Hermitian Hamiltonian
$h:=\eta_+^{1/2}\,H\,\eta_+^{-1/2}$ in the same time $\tau$. As
shown in \cite{prl-2007}, the length of the curve corresponding to
these evolutions in the respective projective Hilbert spaces are
identical. Therefore, they will have the same speed.

Next, we wish to show how by choosing the metric operator we may
adjust the value of $s_{\eta_+}$ and consequently $\tau_{\rm
min}^{(\eta_+)}$. Again, without loss of generality we confine our
attention to the case $N=2$. Let $\{e_1,e_2\}$ denote the standard
basis of $\C^2$, i.e., $e_1:=${\scriptsize$
\left(\begin{array}{c}1\\0\end{array}\right)$}, $e_2:=${\scriptsize$
\left(\begin{array}{c}0\\1\end{array}\right)$}. Then we can
represent any metric operator $\eta_+$ in $\{e_1,e_2\}$ by a
positive-definite matrix of the form
    \be
    \underline{\eta_+}=\left(\begin{array}{cc}
    a & \fb^*\\
    \fb  & c\end{array}\right),
    \label{underline-eta=}
    \ee
where $a,c\in\R$ and $\fb\in\C$. Because $\underline{\eta_+}$ is a
positive-definite matrix,
    \be
    a+c={\rm tr}(\underline{\eta_+})>0,~~~~~
    D:=ac-|\fb|^2=\det(\underline{\eta_+})>0.
    \label{constants=}
    \ee
For states $\lambda_{\psi}$, differing from $\lambda_{e_2}$, we can
use a representative state vector of the form $\psi:=${\scriptsize$
\left(\begin{array}{c}1\\x+iy\end{array}\right)$}. Substituting this
relation in (\ref{ph-Fubini-Study=}) and using (\ref{subs}) and
(\ref{underline-eta=}) we find \cite{prl-2007}
    \be
    ds_{\eta_+}^2=\frac{D(d x^2+d y^2)}{\left[a+2
    (b_1 x+b_2 y)+c( x^2+ y^2)\right]^2},
    \label{prl-8-cor}
    \ee
where $b_1$ and $b_2$ are respectively the real and imaginary parts
of $\fb$, i.e., $\fb=:b_1+i b_2$.

Next, we introduce the angular coordinates $(\varphi,\theta)$ that
are related to $(x,y)$ according to
$x=\tan(\mbox{$\frac{\theta}{2}$}) \cos(\varphi+\beta)$,
$y=\tan(\mbox{$\frac{\theta}{2}$})\sin(\varphi+\beta)$, where
$\beta:=\tan^{-1}(b_2/b_1)$. These coordinates also allow for
treating the state $\lambda_{e_2}$. To see this, first observe that
for $(x,y)\in\R^2$, we have $\varphi\in[0,2\pi)$ and
$\theta\in[0,\pi)$. The state $\lambda_{e_2}$ corresponds to the
point at infinity in the $x$-$y$ plane which we can identify with
$\theta=\pi$. In terms of $(\varphi,\theta)$, (\ref{prl-8-cor})
reads
    \be
    ds_{\eta_+}^2=
    \frac{k_1\:(d\theta^2+\sin^2\theta\,
    d\varphi^2)}{\left[1+k_2\cos\theta+
    k_3\cos\varphi\,\sin\theta\right]^2},
    \label{prl-8-cor-sph}
    \ee
where we have introduced
    $k_1:=\frac{D}{(a+c)^2}=
    \frac{{\det(\underline{\eta_+})}}{
    {\rm tr}(\underline{\eta_+})^2}$,
    $k_2:=\frac{a-c}{a+c}$, and
    $k_3:=\frac{2|\fb|}{a+c}$.
Note that because of (\ref{constants=}) we have $k_1>0$, $-1<k_2<1$
and $0\leq k_3<1$.

If we set $\eta_+=I$ we recover the standard Euclidean inner product
on the Hilbert space. In this case $s^{\eta_+}=s$, $a=c=1$, $\fb=0$,
$k_1=1/4$, $k_2=k_3=0$, and (\ref{prl-8-cor-sph}) becomes $ds^2=
\frac{1}{4}(d\theta^2+\sin^2\theta\,d\varphi^2)$. This is just the
standard metric for a round sphere of unit diameter. In
\cite{prl-2007} we show that $\cP(\pH)$ is related to $\cP(\cH)$ by
an isometry. Therefore, (\ref{prl-8-cor-sph}) also describes a round
sphere of unit diameter, and $(\varphi,\theta)$ are the usual
spherical coordinates.

%As seen from (\ref{prl-8-cor-sph}) the parameter $k_1$
%(alternatively $D$) scales the distance in $\cP(\pH)$. Therefore, by
%adjusting it properly we can make $\tau_{\rm min}^{(\eta_+)}$
%arbitrarily small.

We can use (\ref{ph-H-star}) and (\ref{underline-eta=}), to obtain
the explicit form of the optimal-speed $\eta_+$-pseudo-Hermitian
Hamiltonians for given initial $\lambda_{\psi_I}$ and final
$\lambda_{\psi_F}$ states. We can always perform an invertible
linear (basis) transformation in $\cH$ so that
$\lambda_{\psi_I}=\lambda_{e_1}$. Let $\lambda_{\psi_F}$ be an
arbitrary final state (that is different from $\lambda_{e_1}$). Then
we can take $\psi_I= \mbox{\scriptsize$\left(\begin{array}{c}1\\0
\end{array}\right)$}$, and $\psi_F=\mbox{\scriptsize$\left(\begin{array}{c}
\zeta\\1\end{array}\right)$}$, for some $\zeta\in\C$. In view of
(\ref{subs}) and (\ref{underline-eta=}), we have
$\pbr\psi_I|\psi_I\pkt=a$,
$\pbr\psi_F|\psi_F\pkt=\frac{D+|\xi|^2}{a}$,
$\pbr\psi_I|\psi_F\pkt=a\zeta+\fb^*=:\xi$. Inserting these in
(\ref{ph-cosvartheta=}) and using (\ref{ph-min-time}), we find $\cos
s_{\eta_+}=|\xi|/\sqrt{D+|\xi|^2}$ and
    \be
    \tau_{\rm min}^{(\eta_+)}=\frac{\hbar}{E}\:\cos^{-1}\left(
    |\xi|/\sqrt{D+|\xi|^2}\right).
    \label{explicit-tau-min=}
    \ee

Similarly we employ (\ref{ph-H-star}) and (\ref{underline-eta=}), to
obtain the matrix representation of $H_\star^{(\eta_+)}$ in the
basis $\{e_1,e_2\}$:
    \bea
    \underline{H_\star}^{(\eta_+)}
    &=&\frac{iE\:e^{-i\omega}}{4a\sqrt D}
    \left(\begin{array}{cc}
    -a\fb^* & -(D\:e^{2i\omega}+\fb^{*2})\\
    a^2 & a\fb^*\end{array}\right),
    \label{explicit-H}
    \eea
where $\omega:={\rm arg}(\xi)$, i.e., $e^{i\omega}=\xi/|\xi|$. It is
interesting to see that the minimum travel time and optimal-speed
Hamiltonians are respectively determined by the modulus and the
phase of $\xi$. Note also that the right-hand side of
(\ref{explicit-H}) does not have a unique limit as $\xi\to 0$. This
is because $\xi=0$ corresponds to the case that $\lambda_{\psi_I}$
and $\lambda_{\psi_F}$ are antipodal points of $\cP(\pH)$ that are
connected via an infinity of geodesics with equal length.

Setting $a=c=D=1$ and $\fb=0$ in (\ref{explicit-tau-min=}) and
(\ref{explicit-H}), we find the explicit form of the optimal-speed
Hamiltonian and the minimum travel time in conventional quantum
mechanics: $\underline{H_\star}=\frac{iE}{4} \left(\begin{array}{cc}
0 & -e^{i\omega}\\ e^{-i\omega} & 0\end{array}\right)$,
    \be
     \tau_{\rm min}=\frac{\hbar}{E}\:\cos^{-1}\left(
    |\zeta|/\sqrt{1+|\zeta|^2}\right).
     \label{conventional-1}
     \ee
Note that in this case $\xi=\zeta$ and $e^{i\omega}=\zeta/|\zeta|$.
Comparing (\ref{conventional-1}) with (\ref{explicit-tau-min=}), we
see that by keeping $a$ and $\fb$ fixed, so that $\xi$ is left
unchanged, and decreasing the value of $c$ we can make $D$ as small
as we wish. This in turn reduces the value of $\tau_{\rm
min}^{(\eta_+)}$ below that of $\tau_{\rm min}$. For example, we can
set $a=1$ and $\fb=0$. Then, $D=c$, $\xi=\zeta$, and we find
$\underline{H_\star}^{(\eta_+)}=\frac{iE}{4}
    \left(\begin{array}{cc}
    0 & -\sqrt{c}\:e^{i\omega}\\
    \frac{e^{-i\omega}}{\sqrt c} & 0\end{array}\right)$,
and $\tau_{\rm min}^{(\eta_+)}=\frac{\hbar}{E}\:\cos^{-1}\left(
    |\zeta|/\sqrt{c+|\zeta|^2}\right)$.
For $c<1$, this yields $\tau_{\rm min}^{(\eta_+)}<\tau_{\rm min}$.
As we explained above, this observation does not seem to have any
practical implications, if we use $H_\star^{(\eta_+)}$ to generate a
unitary evolution, i.e., consider the dynamics taking place in
$\cP(\pH)$. If we instead consider the dynamics defined by
$H_\star^{(\eta_+)}$ in $\cP(\cH)$, then the travel time is still
given by (\ref{explicit-tau-min=}) (which can be made smaller than
$\tau_{\rm min}$) but the evolution is non-unitary. This is a
manifestation of the nonexistence of an upper bound on evolution
speed for non-unitary evolutions \cite{preprint}. A more interesting
observation is that one can realize this fact using a
quasi-Hermitian Hamiltonian ($\cP\cT$-symmetric Hamiltonians
considered in \cite{bbj-prl-2007} being special cases) \cite{jones};
\emph{there is no upper bound on the speed of quasi-unitary
evolutions}, \footnote{$U:\cH\to\cH$ is quasi-unitary, if it is
$\eta_+$-pseudo-unitary for some metric operator $\eta_+$, i.e.,
$U^{-1}=\eta_+U^\dagger\eta_+^{-1}$. See A.~Mostafazadeh, J.\ Math.\
Phys., {\bf 45}, 932 (2004).}.

To offer a physical interpretation for this result we first recall
that in quantum mechanics, a physical system is represented by a
Hilbert space-Hamiltonian pair $(\cH,H)$. This representation is
however not unique, for unitary-equivalent Hilbert space-Hamiltonian
pairs describe the same system. If $\cH$ and $\pH$ are Hilbert
spaces that have identical vector space structure but different
inner products (say corresponding to the choices $I$ and $\eta_+$
for their metric operators respectively), one can use a single
Hamiltonian operator to represent two different quantum systems,
e.g., $(\cH,H^{(\eta_+)}_\star)$ and $(\pH,H^{(\eta_+)}_\star)$
represent distinct physical systems with quasi-unitary and unitary
dynamical evolutions, respectively. The above argument shows that
while the evolution speed for the latter system is given by
$E/\hbar$, that of the former can be made arbitrarily large. To
determine whether this observation can have practical applications
requires a more detailed investigation of the role of quasi- and
pseudo-Hermitian Hamiltonians in open quantum systems \cite{JS}.

In summary, we have offered a straightforward derivation of an
expression for the most general time-independent optimal-speed
quasi-Hermitian (in particular Hermitian) Hamiltonians and
established by explicit calculation the metric-dependence of the
minimum travel time and metric-independence of the maximum travel
speed. Our analysis confirms the existence of infinitely fast
quasi-unitary evolutions. These might find applications in areas
such as quantum computation and quantum control,
\cite{Margolus,khaneja}. The derivation of an explicit expression
for the most general optimal-speed quasi-Hermitian Hamiltonian, that
we have reported here, is a necessary step in this direction.

\noindent \textbf{{Acknowledgments}}: This project was supported by
the Scientific and Technological Research Council of Turkey
(T\"UB\.{I}TAK) as a part of the 2007 T\"UB\.{I}TAK Science Award in
Basic Sciences, and by the Turkish Academy of Sciences (T\"UBA).

\ed